\documentclass[11pt]{amsart}
\usepackage{geometry}                
\usepackage{graphicx}
\usepackage{amssymb}
\usepackage{epstopdf}
\usepackage{longtable}
\usepackage{psfrag}
\usepackage{overpic}
\usepackage{color}
\usepackage{caption2}
\usepackage{subfigure}

\title{Heat flows inferred from a Parker's-like formula for stable or quasi-stable continents}
\author{Rong Qiang Wei}
\address{College of Earth Sciences, University of Chinese Academy of Sciences, Beijing, PRC, 100049}
\email{wrq1973@ucas.ac.cn}
\date{}

\begin{document}
\maketitle
\begin{abstract}

Surface heat flow is a key parameter for the geothermal structure, rheology, and hence the dynamics of continents. However, the coverage of heat flow measurements is still poor in many continental areas. By transforming the stable nonlinear heat conduction equation into a Poisson's one, we develop a method to infer surface heat flow for a stable or quasi-stable continent from a Parker's-like formula. This formula provides the relationship between the Fourier transform of surface heat flow and the sum of the Fourier transform of the powers of geometry for the heat production (HP) interface in the continental lithosphere. Once the interface geometry is known, one to three dimensional distribution of the surface heat flow can be calculated accurately by this formula. As a case study, we estimate the three-dimensional surface heat flows for the Ordos geological block and its adjacent areas in China on a $1^\circ \times 1^\circ$ grid based on a simple layered constant HP model. Comparing to the measurements, most relative errors of the heat flows inferred are less than 20\%, showing this method is a favorable way to estimate surface heat flow for stable or quasi-stable continental regions where measurements are rare or absent.

\end{abstract}

{\hspace{2.2em}\small Keywords:}

{\hspace{2.2em}\tiny Heat flow Heat production  Geotherms Stable continental lithosphere the Ordos geological block}

\section{Introduction}\label{intro}

Surface heat flow is an important boundary condition for the thermal modeling of the lithosphere of the continent. It is also important for a better understanding of the heat loss of the Earth, the thermo-mechanical evolution of the continental lithosphere. The direct measurement for the heat flow is often technical, time-consuming and expensive. At present, the heat flow measurements cover only a limited surface of the Earth and distribute heterogeneously, although there are over 38,000 heat flow measurements worldwide (Davies, 2013).  In the continents, Goutorbe et al. (2011) pointed out that only 12 per cent of the continents can be sampled using averages over $1^\circ \times 1^\circ$ windows. If the data quality is taken into account, the number of heat flow values will decrease further.

Some empirical methods were presented to extrapolate heat flow to regions where no measurement covers, except ordinary interpolation and /or extrapolation. For example,  some early authors tried to relate heat flow to the thermotectonic age of the crust and the continental lithosphere (e.g., Vitorello and Pollack, 1980), or to the crustal thickness (e.g., Bodri and Bodri, 1985). However, such empirical relations are argued for their weakness and inadequacy in many areas (e.g., Morgan 1984; Jaupart and Mareschal 2007; Jaupart et al., 2016). 

More methods for estimating the surface heat flow are based on "similar idea". For example, Shapiro and Ritzwoller (2004) extrapolated the existing heat-flow measurements to regions where heat flow measurements are rare or absent, guided by a shear velocity model of the upper mantle. Their basic idea is that regions similar in structure should have similar levels of heat flow. By assuming a similar correlation between heat flow and geology, Davies and Davies (2010) and Davies (2013) evaluated the heat flows for the grids at which there are no measurements, and a global map of surface heat flow was presented on a $2^\circ \times 2^\circ$ equal area grid. Goutorbe et al. (2011) thought that similar geodynamic settings should display similar thermal states,  and they introduced two methods (the best combination and similarity methods) to extrapolate heat flow by transposition of existing measurements. What make their work different from others above is that they include more than 20 set of estimators rather than a single one to represent the variety of geodynamic environments. Finally, two global maps of surface heat flow are presented on a $1^\circ \times 1^\circ$ grids. 

Each of methods above can present a possible (or empirical) distribution for the heat flow at the locations where the measurement did not cover, and reflect the general trends of heat flows, although they will never replace actual measurements of heat flow. However, the heat conduction theory related clearly to the surface heat flow was not taken into account in these methods. In this respect, Maule et al. (2005) developed a method that uses satellite magnetic data to estimate the heat flux underneath the Antarctic ice sheet, based on one dimensional stable linear heat conduction equation.  Results related show their method is a favorable way to estimate the regional scale of spatially varying heat flow.  Here we present a method based on stable (nonlinear) heat conduction theory, to estimate one to three dimensional heat flows for the stable quasi-stable continent. We will describe the relative theory and methodology in the next section. As a case study, the heat flows for areas without measurements in the Ordos geological block and its adjacent areas in China are estimated on $1^\circ \times 1^\circ$ cells.

\section{Theory and Methodology}\label{sec2}
 \subsection{Theory}
 We start from the three-dimensional nonlinear equation of the steady heat conduction,
 
\begin{equation}\label{eq1}
\nabla\cdot[(k(T) \nabla T]+A=0
\end{equation}

where $T=T(x,y,z)$ is the temperature, $k(T)$ the coefficient of thermal conductivity which depends only on $T$, $A=A(x,y,z)$ the radioactive HP.

\begin{figure}[htb]
\setlength{\belowcaptionskip}{0pt}
\centering
\begin{overpic}[scale=0.8]{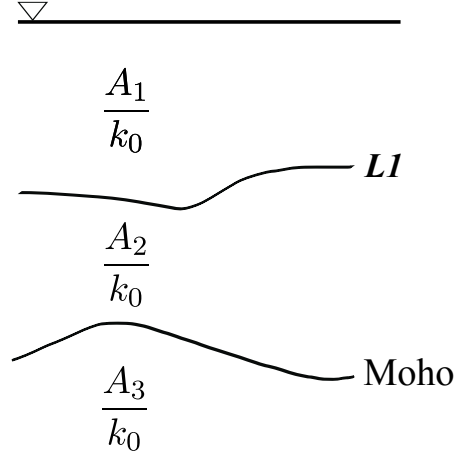}
\end{overpic}
\renewcommand{\figurename}{Fig.}
\caption{The sketch map of the heat production (HP) model for the stable or quasi-stable continental lithosphere. $A_1,A_2,A_3$ are the HP of the upper crust, lower crust, and the lithospheric mantle, respectively. $L1$ and Moho are the HP bottom of the upper crust and lower crust, respectively. The HP bottom of the lithospheric mantle is at infinity downward. $k_0$ is the coefficient of thermal conductivity measured at reference temperature $T_0$.}
\label{fig1}
\end{figure}

By Kirchoff transform,
 
 \begin{equation}\label{eq2}
 U(T)=\int_{_{T_0}}^{T}\frac{k(T')}{k_0}{\rm{d}}T'
 \end{equation}

where $T_0$ is a reference temperature, and $k_0$ the coefficient of thermal conductivity measured at $T_0$.
 
We can transform equation (\ref{eq1}) into (\ref{eq3}) as follows,
 
 \begin{equation}\label{eq3}
\nabla^2 U+\frac{A}{k_0}=0
\end{equation}

Equation (\ref{eq3}) is a Poisson's equation. A special solution $U(\textbf{r})$ of this equation is,

\begin{equation}\label{eq4}
\begin{array}{ll}
U(\textbf{r}) =U(x,y,z) &=\frac{1}{4\pi k_0}\int_V\frac{A(x',y',z') \mbox{d}V}{\vert \textbf{r}-\textbf{r}_0\vert} \\
   &=\frac{1}{4\pi k_0}\int_V\frac{A(x',y',z') \mbox{d}V}{[(x-x')^2+(y-y')^2+(z-z')^2]^{1/2}}
\end{array}
\end{equation}

where $\textbf{r}$ is the radius vector of the field point $(x,y,z)$,  $\textbf{r}_0$ is that of the source point $(x',y',z')$, $V$ the volume of the entire HP distribution.

Further, we can get (\ref{eq5}) from (\ref{eq4}),

\begin{equation}\label{eq5}
\frac{\partial U}{\partial z}=-\frac{1}{4\pi k_0}\int_V\frac{A(x',y',z')(z-z') \mbox{d}V}{[(x-x')^2+(y-y')^2+(z-z')^2]^{3/2}}
\end{equation}

Moreover, according to Kirchoff transform, i.e., equation (\ref{eq2}), we have,

\begin{equation}\label{eq6}
\begin{array}{ll}
 \frac{\partial U}{\partial z} &=\frac{\partial U}{\partial T}\frac{\partial T}{\partial z}\\
  &=\frac{k(T)}{k_0}\frac{\partial T}{\partial z}\\
  &=-\frac{q}{k_0}
  \end{array}
\end{equation}

where $q$ is the heat flow. Equation (\ref{eq6}) shows that the heat flow $q$ is, 

\begin{equation}\label{eq7}
q=-k_0\frac{\partial U}{\partial z}
\end{equation}

At the surface $z=0$, $q=-q_0$ where $q_0$ is the surface heat flow, and we can get,

\begin{equation}\label{eq8}
q_0=k_0\frac{\partial U}{\partial z}\vert_{_{z=0}}
\end{equation}

Equation (\ref{eq5}-\ref{eq8}) mean that surface heat flow can be estimated if the $A(x',y',z')$ is known. 

\subsection{Methodology}

To illustrate the methodology, we construct a three-layered HP model whose bottom is at infinity downward,  as shown in Figure \ref{fig1}. This kind of HP model is typical for continents (e.g., Hasterok and Chapman, 2012).  Like any potential problem,  the surface heat flow may be attributed to $A_1=A_1(x',y', z')$, $A_2=A_2(x',y', z')$, and $A_3=A_3(x',y', z')$ while the geometries for HP interface $L1$ and Moho are known,  or to the geometry of $L1$ and Moho while $A_1$, $A_2$ and $A_3$ are known, or to the two cases above.  At the surface, we can not distinguish them at all only from the measurements of the heat flow. For simplicity, we assume that the second case holds here, and assume further that the HP within the interfaces is constant, i.e., $A_1,A_2,A_2$ are constants.

\begin{figure}[htb]
\setlength{\belowcaptionskip}{0pt}
\centering
\begin{overpic}[scale=0.52]{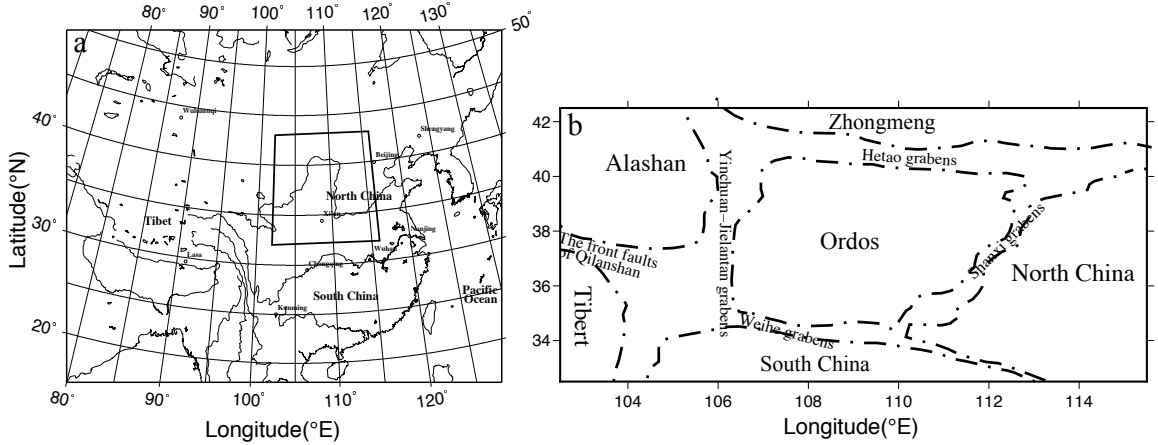}
\end{overpic}
\renewcommand{\figurename}{Fig.}
\caption{(a) Location of the Ordos geological block and its adjacent area in China (the area shown within a box). (b)Map of the study area. The names of the geological blocks are shown in bold fonts. The dash-dotted line are boundary of the geological blocks. The Ordos block is separated by four grabens which consist mainly of faults in different geology age (Zhang et al. 2003). }
\label{fig2}
\end{figure}

\begin{figure}[htb]
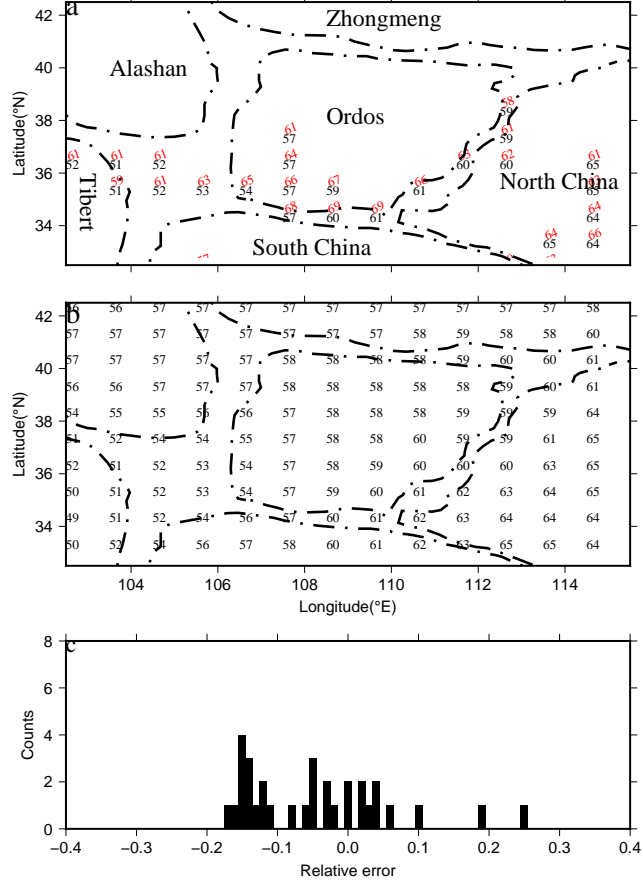

\setlength{\belowcaptionskip}{0pt}
\centering
\begin{overpic}[scale=0.5,bb=15 57 581 751]{Hf_ob_inf_Ordos_min_min_error.eps}
\end{overpic}
\renewcommand{\figurename}{Fig.}
\caption{(a) Comparison of the surface heat flows between the measurements and those inferred by the equation (\ref{eq9}) for the Ordos geological block and its adjacent areas. The red and italic values are heat flow measurements. The black and normal values are those inferred. (b) Surface heat flow estimated on a grid of $1^\circ \times 1^\circ$. (c) The histogram of the relative error.}
\label{fig3}
\end{figure} 

In the potential (gravity or magnetic) problem, there are many methods to obtain anomalies observed (heat flow here) from the geometry of the physical interface. Here we use a method similar to the Paker's approach. Parker (1973) presented an expression that the Fourier transform of the anomalies observed is the sum of the Fourier transforms of the powers of the physical interface causing the anomalies.  Following the Parker's procedures, we consider the heat flow from a layer of HP of constant $A$, whose lower boundary is the plane $z = 0$, and whose upper boundary is defined by the equation $z = h(\textbf{r})$. We have,

\begin{equation}\label{eq9}
F[\frac{\partial U}{\partial z}] =-\frac{A}{2k_0}\exp(-\vert\vec{k}\vert z_0) \sum_{n=1}^{\infty}\frac{\vert\vec{k}\vert ^{n-1}}{n!}F[h^n(\textbf{r})]
\end{equation}

where $F[\cdot]$ denotes the Fourier transform, $\vec{k}$ is the wave number, $h(\textbf{r})$ the depth to the interface (positive downwards) and $z_0$ the depth for the plane to which the observation point is confined. $\exp(-\vert\vec{k}\vert z_0)$ is often taken as the continuation factor. 

Equation (\ref{eq9}) allows us to infer the heat flow caused by the topography $h(\textbf{r})$ of the HP layers.  If $A$, $k_0$, $z_0$, and $h(\textbf{r})$ are given, we can obtain the Fourier transform of the $\frac{\partial U}{\partial z}$, and then the $\frac{\partial U}{\partial z}$. Using  Equation (\ref{eq8}) further, the heat flow at the surface can be estimated.

\begin{figure}[htb]
\setlength{\belowcaptionskip}{0pt}
\centering
\begin{overpic}[scale=0.45]{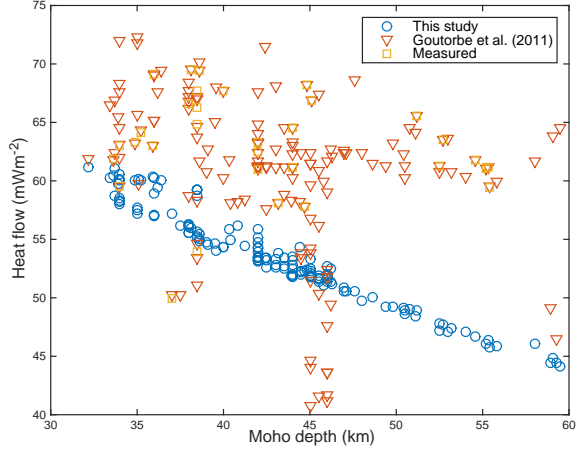}
\end{overpic}
\renewcommand{\figurename}{Fig.}
\caption{Variation of the heat flows with the depths of Moho discontinuities over $1^\circ\times 1^\circ$ grids in Ordos geological block and its adjacent areas. The Moho depths are from CRUST1.0.}
\label{fig4}
\end{figure}

\section{surface heat flow infered for the Ordos geological block and its adjacent areas}\label{sec3}

Because our start point is equation (\ref{eq1}), the methodology to infer the surface heat flow described in section \ref{sec2} is only available to the stable or quasi-stable areas. Here we apply this method to the Ordos geological block and its adjacent areas in China ($102^\circ-116^\circ\rm{E}$,$32^\circ-43^\circ\rm{N}$). As shown in Figure \ref{fig2}, the Ordos geological block is located between the Tibet plateau and the North China plain. There are no active tectonic events such as an uplift of the upper mantle, active Quaternary faults or large earthquakes within the Ordos geological block (Deng et al. 1999).  Therefore this area is stable in tectonics and the methodology in section \ref{sec2} can be used.  To decrease the boundary effects, a broader study area ($100^\circ-120^\circ\rm{E}$,$30^\circ-50^\circ\rm{N}$) to will be adopted.  The calculation is carried out on a grid of $1^\circ \times 1^\circ$. 

\subsection{Data}

The Data required in the calculation are the following,

 1. The HP model and the HP bottom interfaces

We use the simple HP model in Figure \ref{fig1}, i.e., a three-layered one for the Ordos geological block, in which the HPs  are $A_1=1.50\rm{\mu Wm}^{-3}$ in the upper crust, $A_2=0.45\rm{\mu Wm}^{-3}$ in the lower crust, and $A_3=0.02\rm{\mu Wm}^{-3}$ in the lithospheric mantle, respectively (Hasterok and Chapman, 2012; Furlong and Chapman, 2013).  In this model, the depths of $L1$ and the Moho for Ordos geological block and its adjacent areas are from directly the CRUST1.0 (Laske et al., 2013). 

\begin{figure}[htb]
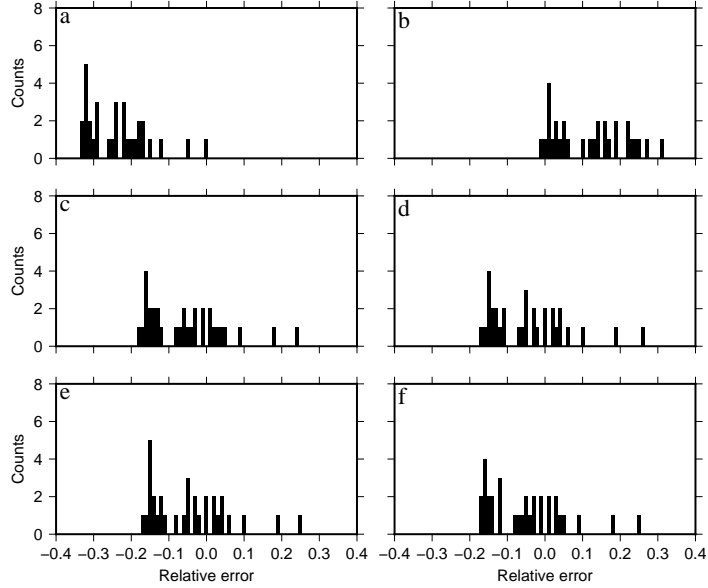

\setlength{\belowcaptionskip}{0pt}
\centering
\begin{overpic}[scale=0.5,angle=0,bb=33 243 564 692]{Effect_A.eps}
\end{overpic}
\renewcommand{\figurename}{Fig.}
\caption{Histogram for relative errors when $A$ of the upper crust, lower crust and the lithospheric mantle have a perturbation of $\pm20$\%, respectively. a. $A$ for the upper crust has a perturbation of $-20$\%; b. the upper crust, 20\%; c. the lower crust, $-20$\%; d. the lower crust, 20\%; e. the lithospheric mantle, $-20$\%; f. the lithospheric mantle, 20\%. }
\label{fig5}
\end{figure}

2. $k_0$ and $z_0$

Here $k_0=3.0\rm{Wm}^{-1}\rm{K} ^{-1}$ (eg., Turcotte and Schubert, 2014). 

$z_0$ is a key parameter in this method. For Ordos geological block, $z_0$ for $L1$ and Moho are 39 km and 43 km, respectively. They are determined from the measurements of the surface heat flows in the least squares sense. These heat flows measured are from the global data set ($1^\circ\times 1^\circ$ grid) (Goutorbe et al. , 2011). 

\subsection{Surface heat flow inferred for the Ordos geological block}\label{hf_inf_ordos}

We take a look at the surface heat flows inferred with equation (\ref{eq9}) for Ordos geological block and its adjacent areas. Figure \ref{fig3}a shows the comparison between the heat flow measurements and those estimated. Figure \ref{fig3}b shows the histogram of the relative errors. It can be seen that the general trend reflected by the heat flow measured (the red and italic) is preserved by those inferred (the black and normal); the maximum relative error is about 25\%; Most of the relative errors are less than 20\%.  In term of the relative error, the inferred heat flows are acceptable, and could be the reasonable estimation for the grids where the heat flow measurement did not cover.  

\begin{figure}[htb]
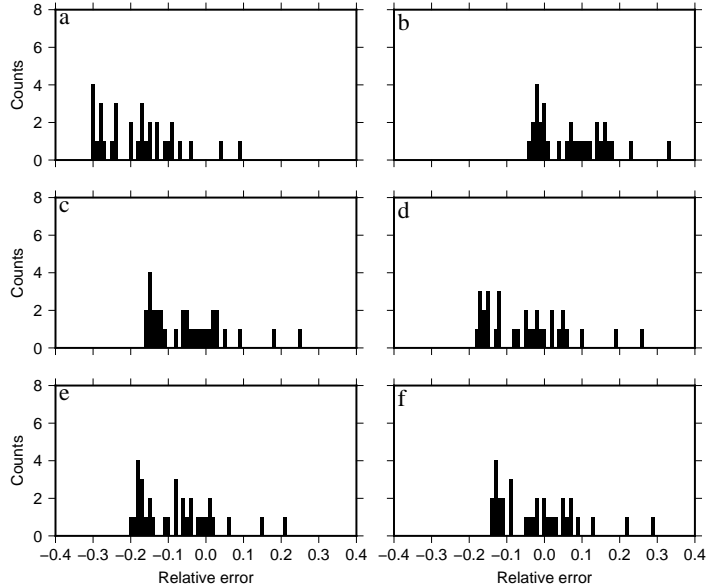

\setlength{\belowcaptionskip}{0pt}
\centering
\begin{overpic}[scale=0.5,angle=0,bb=33 243 564 692]{Effect_jiegou.eps}
\end{overpic}
\renewcommand{\figurename}{Fig.}
\caption{Histogram for relative errors when $z_0$, and depths of $L1$ and Moho have a perturbation, respectively. a. $z_0$ has a perturbation of $-20$\%; b. $z_0$, 20\%; c. depths of $L1$, $-10$\%; d. depths of $L1$, $10$\%; e. depths of Moho, $-10$\%; f. depths of Moho, $10$\%. }
\label{fig6}
\end{figure}

Relatively speaking,  a broad low surface heat flow anomaly is assigned to the Ordos geological block and its adjacent areas.  The main possible reason is that only the contribution from the tomography of the HP interface is taken into account. In fact, the surface heat flow measured contains not only this contribution, but also that from the distribution of the HP itself, even that from local ‘thermotectonic’ setting, and so on. The latter contributions are omitted when we infer the heat flows with equation (\ref{eq9}).  

These heat flows inferred are basically consistent with the tectonic setting of this geological block. However, it can be observed that heat flows inferred change slightly. This largely attributes to the small variations of the $L1$ and Moho beneath the Ordos geological block and its adjacent areas, based on the discussions above.

\subsection{Effect of the parameters of $A$, $z_0$, and $h(\textbf{r})$}

The parameters, such as, radiogenic HP ($A$), $z_0$, and the tomography of the HP interface $h(\textbf{r})$, have influences on the finial heat flow inferred with equation (\ref{eq9}). We estimate the influences by adding a perturbation to these parameters. We also take the Ordos geological block and its adjacent areas as an example.

Figure \ref{fig5} shows the histogram of the relative error when $A$ of the upper crust, lower crust and the lithospheric mantle has a perturbation of $\pm20$\%, respectively. It can be observed from Figure \ref{fig5} that the $A$ of the upper crust has the greatest influences: The absolute value of the maximum relative error is greater than 30\%, but most of absolute values of the relative error are less than 20\% when the $A$ has a reduction of 20\%; When the $A$ has a increase of 20\%, the maximum relative error is greater than 30\%, but most of the relative errors are less than 25\%. It can also be found from Figure \ref{fig5} that $A$ in the lower crust and/or lithospheric mantle has relatively little influences when they have a perturbation of $\pm20$\%: Most of the relative error are less than 20\%.

Figure \ref{fig6}a and b shows the histogram of the relative error when $z_0$ has a perturbation of $\pm20$\%, respectively. It can be observed that  $z_0$ has great influences: Most of absolute values of the relative error are greater than 20\% when $z_0$ has a reduction of 20\%;The maximum is about 30\%. When $z_0$ has a increase of 20\%, the maximum relative error is greater than 30\%, but most of the relative errors are less than 20\%. 

Figure \ref{fig6}c-f shows the histogram of the relative error when the depths of $L1$ and Moho have a change of 10\%, respectively. It can be seen that most of the relative errors are less than 20\%. 

Analyses above show that the great errors to the surface heat flow are from $A$ in the upper crust, $z_0$, and depths of $L1$ or Moho. Because $A$ in the upper crust can be constrained by the measurements of the HP of rocks, and depths of $L1$ or Moho can be constrained by seismological studies, the great uncertainties are from $z_0$. How to determine $z_0$ is important for the method here. On the other hand, it may be improper that a single and uniform $z_0$ is used here to infer the surface heat flow. 

\subsection{Correlation between heat flow and crustal thickness}

Correlation between heat flow and crustal thickness is an age-old problem. However, a clear correlation seems not to be supported by most of the measurements at present. Analysis of global sets over $1^\circ\times 1^\circ$ grids by Jaupart et al. (2016) reveals that the best linear “fit” to the data points has only a correlation coefficient $R=-0.24$. Figure \ref{fig4} shows the scatter plot of heat flow and crustal thickness in the Ordos geological block and its adjacent areas. It can be observed that there is also no a clear correlation between these two data sets when the heat flows are from Goutorbe et al. (2011) and those measured, respectively. The best linear “fit” to the data points has only a correlation coefficient $R=-0.29$ and $R=-0.11$, respectively. This only reveals the existence of a weak negative correlation between surface heat flow and crustal thickness.

For our case (blue circles in Figure \ref{fig4}), there is a clear negative correlation between surface heat flow and crustal thickness with a correlation coefficient $R=-0.97$, which is consistent with the tendency above. However, it should be noted that we cannot use this to try and estimate surface heat flow from crustal thickness.  The reasons are obvious: Our heat flows are inferred based on a layered constant radiogenic HP model (Figure \ref{fig1}), and only the effect of the tomography of the HP interface is taken into account.  We list this result just to test under what conditions there is a clear correlation between surface heat flow and crustal thickness. Even so, it can be seen that this negative correlation above differs from that for a statistically homogeneous crust, in which a positive correlation between heat flow and Moho depth should be expected, because the the total HP, and thus the surface heat flow increases with crustal thickness.

\section{Summary}

Unlike many geophysical features, surface heat flow measurements are unknown on many parts of the Earth. Here we present a method based on the heat conduction theory, to estimate the heat flow for the stable or quasi-stable continental regions where measurements are rare or absent. A Parker's-like formula for calculating the heat flow due to the geometry of the HP interface is derived. This method is rigorous in theory. The case study in the Ordos geological block and its adjacent areas in China, based on a simple layered constant HP model, shows that the maximum relative error to the heat flow measurement is about 25\% and most of the relative errors are less than 20\%. Such inferred heat flows are reasonable estimation from the point of view of the actual measurement for the surface heat flows without measurements. 

In our method, $z_0$, the depth for the plane to which the observation point is confined, has great influences on the inferred heat flows.  Tests from the simple layered HP model above show the maximum relative error may be greater 30\%.  New approaches should be developed to determined this parameter, or constraints must be put on it.

\vspace{10em}

\def\thebibliography#1{
{\Large\bf  References}\list
 {}{\setlength\labelwidth{1.4em}\leftmargin\labelwidth
 \setlength\parsep{0pt}\setlength\itemsep{.3\baselineskip}
 \setlength{\itemindent}{-\leftmargin}
 \usecounter{enumi}}
 \def\newblock{\hskip .11em plus .33em minus -.07em}
 \sloppy
 \sfcode`\.=1000\relax}
\let\endthebibliography=\endlist

\end{document}